\documentclass[aps,pre,twocolumn,groupedaddress]{revtex4}

\usepackage{graphicx}
\usepackage{bm}
\usepackage{amsmath}

\begin{document}

\title{Isotropic-Nematic Transition in Liquid-Crystalline Elastomers:\\
Lattice Model with Quenched Disorder}

\author{Jonathan V. Selinger}
\author{B. R. Ratna}
\affiliation{Center for Bio/Molecular Science and Engineering,
Naval Research Laboratory, Code 6900,\\
4555 Overlook Avenue, SW, Washington, DC 20375}

\date{August 5, 2004}

\begin{abstract}
When liquid-crystalline elastomers pass through the isotropic-nematic
transition, the orientational order parameter and the elastic strain vary
rapidly but smoothly, without the expected first-order discontinuity.  This
broadening of the phase transition is an important issue for applications of
liquid-crystalline elastomers as actuators or artificial muscles.  To understand
this behavior, we develop a lattice model of liquid-crystalline elastomers, with
local directors coupled to a global strain variable.  In this model, we can
consider either random-bond disorder (representing chemical heterogeneity) or
random-field disorder (representing heterogeneous local stresses).  Monte Carlo
simulations show that both types of disorder cause the first-order
isotropic-nematic transition to broaden into a smooth crossover, consistent with
the experiments.  For random-field disorder, the smooth crossover into an
ordered state can be attributed to the long-range elastic interaction.
\end{abstract}


\maketitle

\section{Introduction}

Liquid-crystalline elastomers are complex materials consisting of crosslinked
polymer networks covalently bonded to long, rigid, liquid-crystalline
units~\cite{warner96,terentjev99,warner03}.  Because of this unusual structure,
they combine the elastic properties of rubbers with the anisotropy of liquid
crystals.  Any distortion of the polymer network affects the orientational order
of the liquid crystal, and, likewise, any change in the magnitude or direction
of the orientational order influences the shape of the elastomer.  These
materials have a low-temperature nematic phase, with long-range orientational
order in the liquid-crystalline units, and a high-temperature isotropic phase,
with no long-range orientational order. Near the isotropic-nematic transition, a
small change in temperature induces a large change in the orientational order,
which causes the elastomer to extend or contract substantially.  This thermally
induced extension and contraction enables liquid-crystalline elastomers to be
used as actuators or artificial muscles~\cite{thomsen01,shenoy02,naciri03}.

A key issue for basic and applied research on liquid-crystalline elastomers is
understanding the isotropic-nematic transition.  In conventional liquid crystals
of small molecules, this is a first-order transition, with a discontinuity in
the magnitude of the orientational order as a function of temperature.  By
contrast, experiments on liquid-crystalline elastomers show that both the
orientational order and parameter and the elastic strain vary rapidly but
smoothly across this transition, with no first-order discontinuity~\cite{%
thomsen01,shenoy02,naciri03,schatzle89,kaufhold91,disch94,clarke01}.  Apparently
the experimental behavior is neither a first- nor a second-order transition, but
rather a nonsingular crossover between the isotropic and nematic phases.
Although the transition is sharp enough for applications, it is puzzling from
the theoretical point of view.  We would like to explain the broadening of this
phase transition in liquid-crystalline elastomers, compared with the analogous
transition in conventional liquid crystals.

In a previous paper~\cite{selinger02}, we considered two possible explanations
for this broadening.  The first possibility is that some aligning stress shifts
the transition past a mechanical critical point~\cite{degennes75}, like a
liquid-gas transition at high pressure.  This aligning stress might arise from
an applied tensile stress on the sample, or from an anisotropic internal stress
due to crosslinking an elastomer in the nematic phase.  The second possibility
is that the transition is broadened by heterogeneity in the elastomer.  In
particular, we considered heterogeneity in the local isotropic-nematic
transition temperature.

To assess these two possible explanations, we measured the elastomer strain as a
function of temperature over a range of applied tensile stress.  We used
elastomer samples crosslinked in the nematic phase, which should have a large
anisotropic internal stress imprinted by the crosslinking process, and samples
crosslinked in the isotropic phase, which should not have an anisotropic
internal stress.  By analyzing the experimental data, we found three indications
that the broadening of the phase transition is caused by heterogeneity.  First,
the slope of strain vs.\ temperature at the transition does not depend
sensitively on applied stress, in contrast with the prediction for homogeneous
elastomers.  Second, the broadening occurs even for samples crosslinked in the
isotropic phase.  Third, the data for strain vs.\ temperature could not be fit
well by the predictions of Landau theory for homogeneous elastomers, but could
be fit much better by the homogeneous theory convolved with a heterogeneous
distribution of transition temperatures.

Although our previous study showed the importance of heterogeneity for the
isotropic-nematic transition in liquid-crystalline elastomers, this study still
leaves two open questions.  The first issue is the distribution of local
strains.  The previous theory considered an average over local regions with
different transition temperatures.  At any given temperature, these regions have
different local nematic order parameters and different local strains.  It is not
clear how regions with different local strains can fit together.  The second
issue is the type of heterogeneity.  The previous theory considered a
distribution of the local transition temperature, which could arise from
chemical heterogeneity in an elastomer.  This type of heterogeneity would be
regarded theoretically as \emph{random-bond} disorder.  It is not the only
possible type of heterogeneity.  Another possibility is a distribution of local
stresses, which could arise from local orientational order in different
directions at the time of crosslinking.  This possibility would be regarded
theoretically as \emph{random-field} disorder.  Several recent papers have
considered random-field disorder in liquid-crystalline
elastomers~\cite{fridrikh97,yu98,fridrikh99,yu99,uchida00,xing03}.  These
studies have shown that random fields strongly affect mechanical properties and
correlation functions in the low-temperature nematic phase.  However, they have
not made predictions for the effects of random-bond or random-field disorder on
the temperature-dependent isotropic-nematic transition.

\begin{figure}
\includegraphics[clip,width=3.375in]{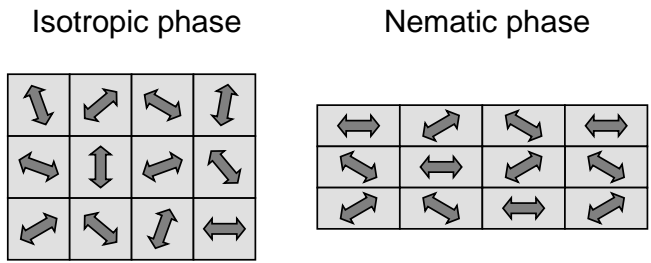}%
\caption{Schematic illustration of the lattice model for the isotropic-nematic
transition in liquid-crystalline elastomers.  In the isotropic phase, the
directors are disordered, and there is no strain.  In the nematic phase, the
directors are ordered along one axis, and the material is extended with strain
along that axis.}
\end{figure}

The purpose of this paper is to develop a lattice model for liquid-crystalline
elastomers, which addresses these theoretical questions about the
isotropic-nematic transition.  In this model, we explicitly consider both
orientational order and elastic strain, as shown in Fig.~1.  Orientational order
is described by a nematic director, which is defined on each site of a
three-dimensional (3D) lattice, as in the Lebwohl-Lasher model of liquid
crystals~\cite{lebwohl72}.  By contrast, elastic strain is defined by a single
global lattice distortion variable.  Because the strain is assumed to be
uniform, there is no problem of fitting together regions with different strains.
In this model, we can consider either random-bond or random-field disorder.
Random-bond disorder enters the model as a variation in the strength of the
local coupling constant between local directors on neighboring lattice sites,
which implies a variation in the local transition temperature.  By comparison,
random-field disorder enters the model as a variation in the direction of an
aligning field that acts on each local director, which implies a local stress on
the elastomer.  Both types of disorder are \emph{quenched}, meaning that they
are fixed and cannot evolve toward equilibrium.

Our study leads to four main results.  First, we derive a new theoretical
formalism for liquid-crystalline elastomers, which translates the
Warner-Terentjev neoclassical rubber
elasticity~\cite{warner96,terentjev99,warner03} into a lattice Hamiltonian for
interacting directors and strain.  Second, we use
Monte Carlo simulations to determine the orientational order parameter and
elastic strain of a homogeneous elastomer as a function of temperature and
applied stress.  These simulations show that the homogeneous elastomer has a
mechanical critical point, as expected from Landau theory.  Third, we simulate
an elastomer with random-bond disorder, and find a broadening of the
isotropic-nematic transition in both the orientational order and the strain.
This result is consistent with experimental data for liquid-crystalline
elastomers.  Fourth, we simulate an elastomer with random-field disorder, and
find that the transition is also broadened in this case, consistent with
experiments, provided that the random field strength is in the right range.  The
result for the random-field system is surprising, because random fields
generally destroy long-range order rather than broadening a transition to an
ordered state.  We attribute this result to an effective long-range interaction
mediated by the elastic strain.

The plan of this paper is as follows.  In Sec.~II we work out the theoretical
formalism, leading to an explicit lattice Hamiltonian that can be simulated.  In
Sec.~III we present the simulations for homogeneous, random-bond, and
random-field elastomers, and give numerical results for the isotropic-nematic
transition in each case.  In Sec.~IV we discuss these numerical results, and
compare them with experiments and with other theoretical studies of quenched
disorder.

\section{Model}

In order to simulate the isotropic-nematic transition in liquid-crystalline
elastomers, we need a mathematical model for the interacting orientational and
elastic degrees of freedom.  In this model, we define a local nematic director
$\bm{n}_i$ on each site $i$ of a 3D cubic lattice.  As in a conventional liquid
crystal, the director $\bm{n}_i$ is equivalent to $-\bm{n}_i$.  The directors
interact with a global lattice distortion tensor $\bm{\lambda}$, which
represents the overall shape of the sample.  A schematic view of the directors
and lattice distortion is shown in Fig.~1.

The Hamiltonian for this lattice model can be written as
\begin{equation}
F=\sum_{\langle i,j\rangle} F_\text{interaction} (\bm{n}_i,\bm{n}_j)
+\sum_i F_\text{elastic} (\bm{n}_i,\bm{\lambda}).
\label{hamiltonian}
\end{equation}
The first term in Eq.~(\ref{hamiltonian}) is an interaction that favors
alignment of the directors on neighboring lattice sites $i$ and $j$.  As in the
Lebwohl-Lasher model, this interaction can be written explicitly as
\begin{equation}
F_\text{interaction} (\bm{n}_i,\bm{n}_j) = -J_{ij} (\bm{n}_i\cdot\bm{n}_j)^2,
\label{interaction}
\end{equation}
where $J_{ij}>0$ is the local bond strength, which may be either uniform or
disordered.  The second term in Eq.~(\ref{hamiltonian}) is an elastic term that
couples the director orientation at lattice site $i$ with the shape of the
polymer chains, which are determined by the overall lattice distortion tensor
$\bm{\lambda}$.  This term should favor alignment of the directors along an
orientation determined by the lattice distortion, and, conversely, favor a
lattice distortion in an orientation determined by the directors.

To develop an explicit expression for this elastic term, we use an argument
based on the neoclassical rubber elasticity of Warner and
Terentjev~\cite{warner96,terentjev99,warner03}.  In their theory, they derive
the general formula
\begin{equation}
F_\text{elastic} = \frac{\mu}{2} \left[\text{Tr}
\left(\bm{\ell}_0\cdot\bm{\lambda}^T\cdot\bm{\ell}^{-1}\cdot\bm{\lambda}\right)
+\ln\left(\frac{\det\bm{\ell}}{a^3}\right)\right],
\label{trace}
\end{equation}
where $\mu$ is the shear modulus, $\bm{\lambda}$ is the lattice distortion
tensor, $\bm{\ell}$ is the shape tensor of the polymer chains, $\bm{\ell}_0$ is
the shape tensor at the time of crosslinking, and $a$ is the average polymer
step length.  They further work out a specific expression for the shape tensor
in the freely jointed chain model.  That specific expression is not appropriate
for a lattice Hamiltonian, because it is based on a \emph{global} average over a
system with \emph{imperfect} nematic order.  By contrast, in a lattice model,
there is a director on each lattice site, which represents \emph{perfect}
\emph{local} orientational order along $\bm{n}_i$ at site $i$.  Any imperfect
long-range order must emerge spontaneously from simulations of the interacting
system.  Hence, we should construct a local shape tensor $\bm{\ell}_i$ for site
$i$, derived from the director $\bm{n}_i$, which will enter into the trace
formula of Eq.~(\ref{trace}).

To construct the local shape tensor $\bm{\ell}_i$, we first consider a
coordinate system aligned along the director $\bm{n}_i$, in which $\bm{\ell}_i$
is diagonal.  In that coordinate system, we have
\begin{equation}
\bm{\ell}_{i}^{-1}=\left(
\begin{array}{ccc}
\ell_{\bot}^{-1} & 0                & 0              \\
0                & \ell_{\bot}^{-1} & 0              \\
0                & 0                & \ell_{\|}^{-1}
\end{array}
\right),
\end{equation}
where $\ell_{\bot}$ and $\ell_{\|}$ are the anisotropic polymer step lengths
favored by the local mesogenic unit.  In a general coordinate system, the shape
tensor components become
\begin{equation}
\ell_{i,\alpha\beta}^{-1}=\ell_{\bot}^{-1}\delta_{\alpha\beta}+
\left(\ell_{\|}^{-1}-\ell_{\bot}^{-1}\right) n_{i,\alpha} n_{i,\beta}.
\label{shapetensor}
\end{equation}

A similar argument gives an explicit expression for the lattice distortion
tensor $\bm{\lambda}$.  Suppose the lattice is uniformly strained along the axis
$\bm{m}$.  In a coordinate system aligned with this strain axis, we have
\begin{equation}
\bm{\lambda}=\left(
\begin{array}{ccc}
\lambda^{-1/2} & 0              & 0       \\
0              & \lambda^{-1/2} & 0       \\
0              & 0              & \lambda
\end{array}
\right),
\end{equation}
where $\lambda$ is the distortion factor, i.e. the strained length normalized by
the original length of the sample, which is related to the strain $e$ by
$\lambda=1+e$. In a general coordinate system, the distortion tensor components
become
\begin{equation}
\lambda_{\alpha\beta}=\lambda^{-1/2}\delta_{\alpha\beta}+
\left(\lambda-\lambda^{-1/2}\right) m_{\alpha} m_{\beta}.
\label{distortiontensor}
\end{equation}

The third tensor required for neoclassical rubber elasticity is the shape tensor
$\bm{\ell}_0$ at the time of crosslinking.  For now, suppose the system is
crosslinked in a totally disordered state, with no long-range or even local
orientational order.  In that case, a reasonable model for $\bm{\ell}_0^{-1}$ is
the isotropic average of $\bm{\ell}^{-1}$.  This average gives the tensor
components
\begin{equation}
\ell_{0,\alpha\beta}^{-1}=a^{-1}\delta_{\alpha\beta},
\label{originalshapetensor}
\end{equation}
where $a^{-1}=(2\ell_{\bot}^{-1}+\ell_{\|}^{-1})/3$.

To determine the elastic term in the lattice Hamiltonian, we substitute the
tensor expressions (\ref{shapetensor}), (\ref{distortiontensor}), and
(\ref{originalshapetensor}) into the general formula of Eq.~(\ref{trace}).  In
this substitution, we note that $\det\bm{\ell}$ is constant because $\bm{\ell}$
represents perfect local orientational order at a specific lattice site.  Hence,
the determinant term adds an unimportant constant to the Hamiltonian, and we can
neglect it.  After some algebra, the trace term leads to
\begin{eqnarray}
\label{elastic}
\lefteqn{F_\text{elastic} (\bm{n}_i,\bm{\lambda}) =} \\
&& \frac{\mu}{2} \left[
\left(\lambda^2 + 2\lambda^{-1}\right)
-\gamma\left(\lambda^2 - \lambda^{-1}\right)
\left(\frac{3}{2}\left(\bm{m}\cdot\bm{n}_i\right)^2 - \frac{1}{2}\right)
\right]. \nonumber
\end{eqnarray}
In this expression, the first term is the classical elastic free energy for
conventional isotropic elastomers, and the second term represents the anisotropy
of liquid-crystalline elastomers.  As expected, the second term shows a coupling
between the elastic strain and the director orientation.  When the lattice is
strained, each local director $\bm{n}_i$ tends to align along the strain axis
$\bm{m}$, with an aligning potential that increases as the distortion $\lambda$
increases above $1$.  Conversely, when the directors are aligned, the lattice
tends to extend along the average director.  The strength of the coupling
depends on the parameter
\begin{equation}
\gamma=
\frac{2\ell_{\bot}^{-1}-2\ell_{\|}^{-1}}{2\ell_{\bot}^{-1}+\ell_{\|}^{-1}},
\end{equation}
which represents the difference in polymer step lengths parallel and
perpendicular to the local director.  This parameter expresses the anisotropy of
the local mesogenic units, and controls how the local director interacts with
the strain.  It ranges from 0 (in the isotropic case $\ell_{\|}=\ell_{\bot}$) to
1 (in the maximally anisotropic limit $\ell_{\|}\gg\ell_{\bot}$).

We must now consider the possibility of symmetry-breaking fields acting on the
elastomer.  Symmetry-breaking fields can arise from two possible sources.  The
simplest possibility is a uniform stress $\sigma$ applied to the elastomer.
Such a stress couples to the strain $e$, or equivalently to the distortion
$\lambda=1+e$, and gives an additional contribution to the Hamiltonian of
$-\sigma\lambda$ for each lattice site.  A more subtle possibility is a
symmetry-breaking field quenched into the local shape tensor $\bm{\ell}_0$ at
the time of crosslinking.  If the system has long-range order at the time of
crosslinking, then $\bm{\ell}_0$ is anisotropic with a single principal axis
$\bm{n}_0$ at all lattice sites.  If the system has short-range order at the
time of crosslinking, then $\bm{\ell}^i_0$ is anisotropic with a different
principal axis $\bm{n}^i_0$ at each lattice site $i$.

In principle, we can incorporate the long- or short-range anisotropy of
$\bm{\ell}^i_0$ into the model by writing a general expression for this tensor
and substituting it into the trace formula of Eq.~(\ref{trace}).  The detailed
calculation is not algebraically tractable, but by symmetry we can see that the
tensor components $\ell^i_{0,\alpha\beta}$ must involve a combination of the
isotropic tensor $\delta_{\alpha\beta}$ and the anisotropic tensor
$n^i_{0,\alpha}n^i_{0,\beta}$ at site $i$.  The anisotropic term acts as an
effective field on the director $\bm{n}_i$, with a coupling of the form
$-(\bm{h}_i\cdot\bm{n}_i)^2$.  The direction of $\bm{h}_i$ is the local
quenched-in axis $\bm{n}^i_0$, and the magnitude of $\bm{h}_i$ scales with the
magnitude of the local quenched-in nematic order.

By combining Eq.~(\ref{interaction}), Eq.~(\ref{elastic}), and the effects of
symmetry-breaking fields, we obtain the final expression for the lattice
Hamiltonian
\begin{eqnarray}
F & = & -\sum_{\langle i,j\rangle} J_{ij} (\bm{n}_i\cdot\bm{n}_j)^2
+\sum_i \biggl[\frac{\mu}{2} \left(\lambda^2 + 2\lambda^{-1}\right)
\nonumber \\
&& \ \ \ \ -\frac{\mu\gamma}{2} \left(\lambda^2 - \lambda^{-1}\right)
\left(\frac{3}{2}\left(\bm{m}\cdot\bm{n}_i\right)^2 - \frac{1}{2}\right)
\nonumber \\
&& \ \ \ \ -\sigma\lambda-\left(\bm{h}_i\cdot\bm{n}_i\right)^2 \biggr].
\label{finalhamiltonian}
\end{eqnarray}
In this Hamiltonian, the statistical variables are the director $\bm{n}_i$ on
each lattice site $i$ and the overall lattice distortion $\lambda$.  We can
assume that the distortion axis $\bm{m}$ is aligned with the principal axis of
the director distribution.

This model of Eq.~(\ref{finalhamiltonian}) can describe uniform elastomers or
elastomers with quenched random-bond or random-field disorder.  Uniform
elastomers are represented by a bond strength $J_{ij}=J$ independent of
position, and by a local field $\bm{h}_i=0$.  In this case, the system can have
an isotropic-nematic transition with a transition temperature $T_{IN}$ that
depends on $J$.  Random-bond elastomers are represented by a bond strength
$J_{ij}$ that depends on the position of the lattice sites $i$ and $j$.  We can
regard this variation in the bond strength as a variation in the local $T_{IN}$,
which could be caused by chemical heterogeneity in the elastomer.  Random-field
elastomers are represented by a local field $\bm{h}_i$ that varies randomly with
position.  This variation models randomness in the local orientational order at
the time of crosslinking, which gives heterogeneous local stresses on the
elastomer.  We can now perform simulations to study the isotropic-nematic
transition in each scenario.  These simulations are presented in the following
section.

\section{Simulations}

We simulate the model of Eq.~(\ref{finalhamiltonian}) using the Monte Carlo
method.  We run the simulations on a 3D cubic lattice with periodic boundary
conditions.  We use a 3D rather than a 2D lattice, even though the simulations
take longer in 3D, in order to avoid a 2D Kosterlitz-Thouless
transition~\cite{kosterlitz73}.  We normally use a lattice of size
$36\times36\times36$.  However, we have run a limited number of simulations on a
larger lattice of size $48\times48\times48$, for the uniform, random-bond, and
random-field cases, and the results are generally consistent.  In the
simulations, we take the uniform or average value of the bond strength $J_{ij}$
to be 1, and the shear modulus $\mu$ to be 1.  These parameters define a scale
for the temperature.  We let the anisotropy parameter $\gamma$ have its maximum
value of 1, in order to see the greatest coupling between orientational order
and elastic distortion.

In each Monte Carlo step of the simulations, we attempt one local director
rotation per lattice site and one change in the overall elastic distortion
$\lambda$.  At each temperature, we equilibrate for 3000 Monte Carlo steps, and
then collect data for 2000 Monte Carlo steps. This number of steps is sufficient
to reach equilibrium at all temperatures except for certain cases of hysteresis,
which are discussed below.  From the numerical data, we extract two parameters
as functions of temperature $T$: the elastic distortion $\lambda(T)$ and the
orientational order parameter $S(T)$, which is defined as the largest eigenvalue
of the tensor $Q_{\alpha\beta}=\langle\frac{3}{2}n_{i,\alpha}n_{i,\beta}
-\frac{1}{2}\delta_{\alpha\beta}\rangle$, averaged over lattice sites $i$.  That
parameter represents the degree of ordering of the local directors along an
average axis.

We begin with the local directors in a disordered configuration, and then cycle
the temperature downward and back upward, using the ending configuration at one
temperature as the starting point for the next.  This temperature cycling mimics
the procedure in typical experiments, and provides an explicit test for
hysteresis.  For most parameter sets, we vary the temperature from 0.9 to 0.8
and back in steps of 0.05, for a total of 41 runs over the temperature range.
Each temperature cycle requires approximately 48 hours on a single processor of
the Huinalu linux supercluster at the Maui High Performance Computing Center.

Because the temperature cycle gives two runs at each temperature (except the
lowest), we can assess whether each run shows a stable, metastable, or unstable
state.  To test for unstable states, we check whether the order parameter $S$
has stabilized by fitting it as a linear function of the Monte Carlo step number
over the final 2000 steps at each temperature.  We identify a state as unstable
and remove it from our results if the absolute value of the slope exceeds a
threshold.  In practice, we find that the threshold of $3.2\times10^{-5}$
eliminates at most two runs from each temperature cycle, one on cooling and one
on heating.  To test for stability vs.\ metastability, we compare the values of
$S$ for cooling and heating runs at the same temperature.  If these values are
within six standard deviations of each other, we assume they represent the same
stable state, so we average the two runs to obtain one data point with improved
statistics.  If not, we assume that one state is stable and the other
metastable, so we report both in our results.

\subsection{Uniform Elastomers}

\begin{figure}
\includegraphics[clip,width=3.375in]{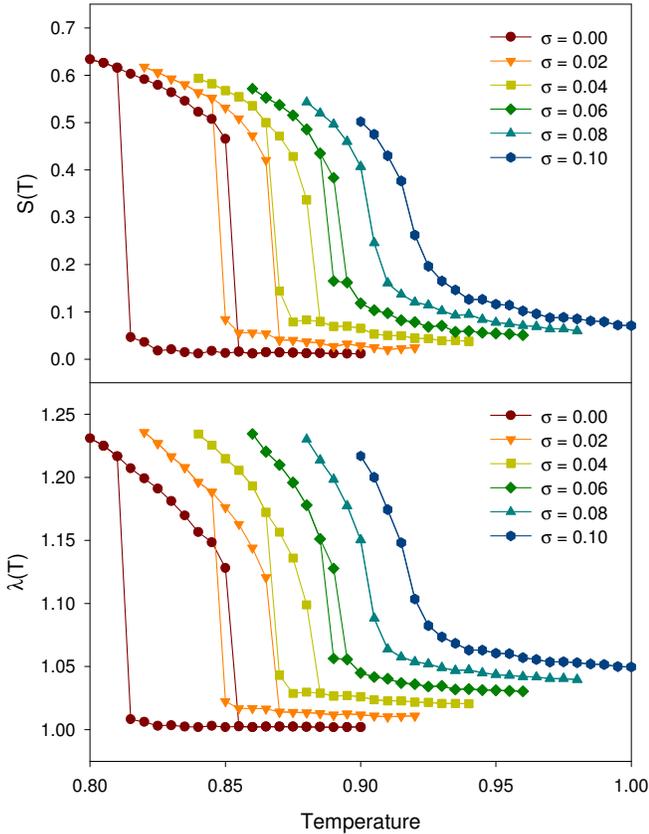}%
\caption{(Color online) Plots of the orientational order parameter $S(T)$ and
the elastic distortion $\lambda(T)$ for a homogeneous elastomer under several
values of the applied uniform stress $\sigma$.  At zero stress, both plots show
a first-order isotropic-nematic transition with a large hysteresis.  As the
stress increases, the first-order discontinuity decreases and then vanishes as
the system passes through a mechanical critical point.  Temperature is in units
of the bond strength $J$.}
\end{figure}

For an initial series of simulations, we consider uniform elastomers, with no
randomness in the bonds (all $J_{ij}=1$) and no random fields (all
$\bm{h}_i=0$).  The numerical results for these simulations are shown in Fig.~2.
For zero applied stress $\sigma$, the system has a first-order transition with
hysteresis between the high-temperature isotropic phase and the low-temperature
nematic phase.  On cooling, the orientational order parameter $S(T)$ jumps from
0.05 to 0.62, and the lattice distortion $\lambda(T)$ jumps from 1.01 to 1.22
(1\% to 22\% strain), at a scaled temperature of 0.81.  On heating, $S(T)$ jumps
from 0.47 to 0.02, and $\lambda(T)$ jumps from 1.13 to 1.00 (13\% to 0\%
strain), at a scaled temperature of 0.85.  The large jumps in both of these
order parameters, and the width of the hysteresis region, show the strong
first-order character of the transition.

When a symmetry-breaking stress is applied to the uniform elastomer, the phase
transition changes drastically.  An applied stress increases both order
parameters $S(T)$ and $\lambda(T)$ for all temperatures.  As the stress
becomes larger, the transition temperature increases, the first-order jumps in
the order parameters decrease, and the hysteresis region becomes narrower.  At a
critical value of the stress between 0.06 and 0.08, the first-order jumps vanish
and the hysteresis goes away.  Beyond that stress, the system shows a smooth
supercritical evolution between the high-temperature disordered limit and the
low-temperature ordered limit.  As the stress continues to increase, this
supercritical evolution becomes increasingly broad.  This trend with increasing
stress is consistent with the prediction of de Gennes based on symmetry
considerations~\cite{degennes75}.  It is analogous to the critical point in the
liquid-gas transition under high pressure.

The simulations show that the orientational order parameter $S(T)$ and the
elastic strain $e(T)=\lambda(T)-1$ have roughly the same dependence on both
temperature and applied stress.  The linear scaling between $S(T)$ and $e(T)$ is
consistent with the prediction based on symmetry considerations.

We note that the smooth evolution of $S(T)$ and $e(T)$ beyond the mechanical
critical point agrees with experiments on the isotropic-nematic transition in
liquid-crystalline elastomers.  However, as discussed in the Introduction, our
previous paper found experimental indications that the isotropic-nematic
transition is smooth even if an elastomer is \emph{not} under a supercritical
stress~\cite{selinger02}. For that reason, we need to look for other mechanisms
to broaden this transition.  Hence, we consider random-bond and random-field
disorder in the following sections.

\subsection{Random Bonds}

\begin{figure}
\includegraphics[clip,width=3.375in]{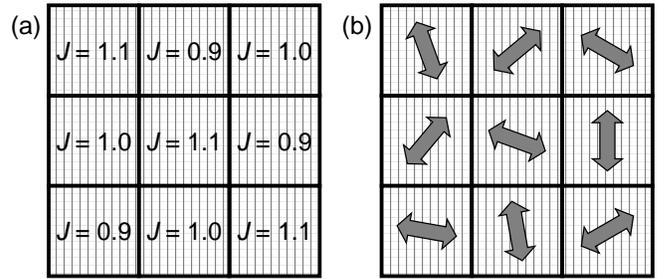}%
\caption{Schematic view of the block structure in the simulations with quenched
disorder.  (a)~Random bond strength.  (b)~Random field orientations.}
\end{figure}

To simulate disordered systems, we first consider elastomers with variations in
the local bond strength.  We let the parameters $J_{ij}$ of
Eq.~(\ref{finalhamiltonian}) be quenched random variables, which are fixed at
the beginning of the simulation and do not change in response to the
statistical evolution of the directors and the lattice distortion.  We suppose
that the variation of $J_{ij}$ occurs in blocks, as shown in Fig.~3(a).  Within
each block, $J_{ij}$ has a uniform value from a Gaussian distribution with mean
1.  From block to block, there are no correlations in $J_{ij}$.  Both the width
of the Gaussian distribution and the size of the blocks are parameters for the
model, which are discussed below.  Because the isotropic-nematic transition
temperature depends on the bond strength, this model represents an elastomer
with blocks of different local transition temperature $T_{IN}$.  This could
occur if the elastomer is chemically heterogeneous, with different compositions
in different local regions.

\begin{figure}
\includegraphics[clip,width=3.375in]{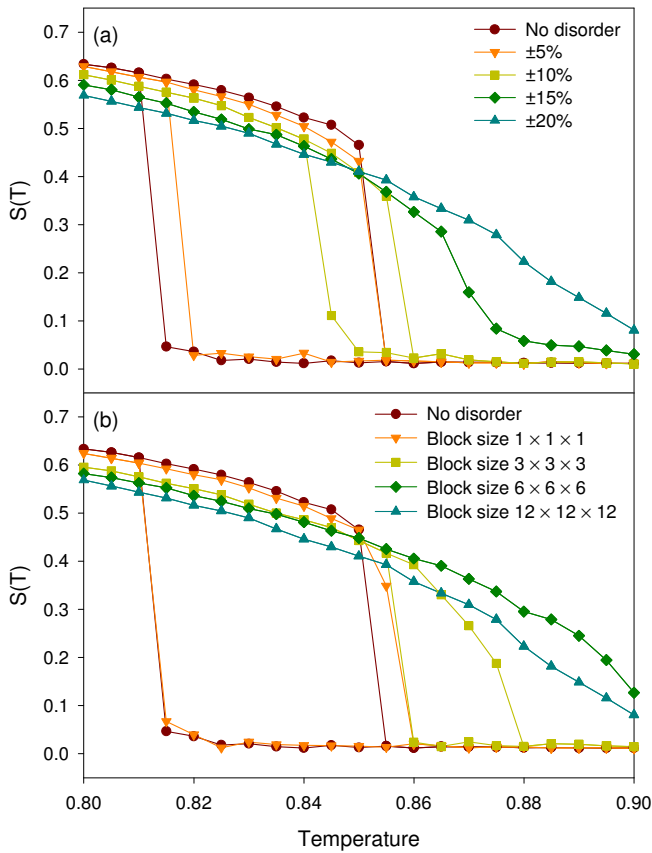}%
\caption{(Color online) Orientational order parameter $S(T)$ for simulations of
elastomers with random-bond disorder.  (a)~Varying magnitude of disorder, with
fixed block size $12\times12\times12$.  (b)~Varying block size of disorder, with
fixed magnitude $\pm20\%$.  As the disorder magnitude and block size increase,
the transition is broadened into a smooth crossover between the isotropic and
nematic limits.  Temperature is in units of the average bond strength $J_{ij}$.}
\end{figure}

In Fig.~4(a), we show the results for varying the magnitude of the disorder,
i.e.\ the standard deviation of the Gaussian distribution of $J_{ij}$, for the
fixed block size $12\times12\times12$.  To save space, we show only the plots of
the orientational order parameter $S(T)$, because the corresponding plots of the
lattice distortion $\lambda(T)$ look quite similar.  For weak disorder of
$\pm5\%$, the system has a strong first-order isotropic-nematic transition,
which is almost identical to the result for no disorder.  Clearly the
orientational order can average over this weak disorder strength.  For a larger
disorder of $\pm10\%$, there is still a first-order transition, but the
first-order discontinuity in $S(T)$ is smaller, and the width of the hysteresis
region is greatly reduced.  When the disorder reaches $\pm15\%$, there is no
longer a first-order discontinuity nor a hysteresis loop.  Instead, the material
evolves rapidly but smoothly between the isotropic and nematic limits as a
function of temperature.  For an even larger disorder of $\pm20\%$, the
transition becomes even broader, with a reduced slope in $S(T)$.

Figure~4(b) presents the results of varying the block size for a fixed disorder
magnitude of $\pm20\%$.  For the large block size $12\times12\times12$ discussed
above, or even for block size $6\times6\times6$, the random-bond disorder causes
a smooth evolution between the isotropic and nematic states.  However, for a
reduced block size of $3\times3\times3$, there is a first-order
isotropic-nematic transition, with a small first-order discontinuity and small
hysteresis.  If the block size is reduced to $1\times1\times1$, i.e. each block
is a single lattice site, then the system has a strong first-order transition.
Although the disorder strength is quite large, the result for block size
$1\times1\times1$ is almost identical to the result for no disorder.  Hence,
reducing the block size is effectively equivalent to reducing the magnitude of
disorder.  The orientational order can average over small blocks of large
disorder, just as it averages over large blocks of small disorder.

The results of this section show that random-bond disorder can change the nature
of the isotropic-nematic transition in liquid-crystalline elastomers, provided
that the magnitude \emph{and} length scale of the disorder are sufficiently
large.  If those conditions are satisfied, then the orientational order
parameter $S(T)$ undergoes a continuous change from a low value in the
high-temperature isotropic limit to a large value in the low-temperature nematic
limit.  The lattice distortion $\lambda(T)$ goes through a corresponding smooth
evolution.  Thus, this type of disorder provides one mechanism to explain the
experimental results.

\subsection{Random Fields}

As an alternative to random bonds, quenched disorder might affect
liquid-crystalline elastomers through random fields coupling to the local
directors.  To simulate random-field effects, we let the fields $\bm{h}_i$ of
Eq.~(\ref{finalhamiltonian}) be quenched random variables, and let the bond
strengths $J_{ij}$ be fixed at 1.  We suppose that the fields $\bm{h}_i$ have a
fixed magnitude $h$ and random orientation.  As in the random-bond case, we
suppose that the randomness occurs in blocks, as shown in Fig.~3(b).  The random
orientation is uniform at every site within a block, and it has no correlations
from block to block.  The magnitude $h$ of the random field and the size of the
blocks are thus two parameters for this model.  Note that this model represents
an elastomer with different preferred orientations of the local director in
different blocks.  This could occur if the crosslinking process quenches
heterogeneous local stresses into the polymer network.

\begin{figure}
\includegraphics[clip,width=3.375in]{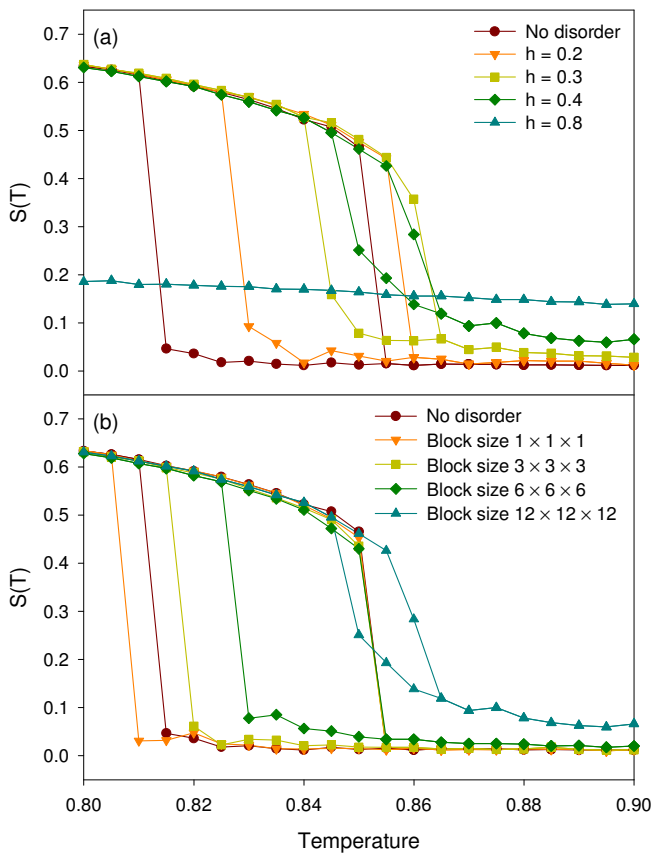}%
\caption{(Color online) Orientational order parameter $S(T)$ for simulations of
elastomers with random-field disorder.  (a)~Varying magnitude of random field,
with fixed block size $12\times12\times12$.  (b)~Varying block size, with fixed
random-field strength $h=0.4$.  As the random-field strength and block size
increase, the transition is broadened and then the ordered nematic phase is
destroyed.  Temperature is in units of the bond strength $J$.}
\end{figure}

Simulation results for several values of the random-field strength $h$ at fixed
block size $12\times12\times12$ are presented in Fig.~5(a).  For a small random
field $h=0.2$, the system has a first-order isotropic-nematic transition, which
is fairly close to the result for no disorder.  For a slightly larger random
field $h=0.3$, the magnitude of the first-order discontinuity and the width of
the hysteresis region are both reduced.  For $h=0.4$, the first-order transition
is much weaker, and the system is close to the smooth crossover between the
isotropic and nematic phases seen in the previous two sections.  However, for a
larger random field $h=0.8$, the behavior is quite different.  Instead of
broadening the isotropic-nematic transition, the random field simply destroys
the long-range nematic order in an athermal way.  In that high-field limit, each
local director is just aligned with its local random field, giving a slight
residual order parameter $S(T)$ that is approximately independent of
temperature.

Figure~5(b) shows the simulation results for several values of the block size at
the fixed random-field strength $h=0.4$.  As in the random-bond case, varying
the block size has the same effect as varying the random-field strength.  For
small block size, the system has a first-order transition that is very close to
the result for no disorder.  This behavior shows that the orientational order
averages over small blocks of strong random field.  For larger block size, the
phase transition gradually changes toward a smooth crossover between the
isotropic and nematic limits.  Analogous simulations for $h=0.8$ (not shown)
demonstrate that increasing the block size takes the system from a strong
first-order transition toward a limit in which orientational order is destroyed
for all temperature.

Overall, the simulations presented in this section show that random fields can
have two distinct effects on the isotropic-nematic transition in
liquid-crystalline elastomers.  Low random-field disorder broadens the
isotropic-nematic transition, but high random-field disorder destroys the
nematic phase in a temperature-independent way.  The results for low
random-field disorder are consistent with experiments on liquid-crystalline
elastomers, but the results for high random-field disorder show a very different
type of behavior.

\section{Discussion}

In this paper, we have investigated three possibile mechanisms to explain the
broadening of the isotropic-nematic transition in liquid-crystalline elastomers.
The first mechanism is a stress that couples to the strain and hence to the
nematic order.  Our simulations show that a small stress reduces the first-order
discontinuity in the isotropic-nematic transition, and a critical stress causes
this discontinuity to vanish.  Beyond that point, the elastomer has a smooth
crossover between the isotropic and nematic limits as a function of temperature.
Hence, a supercritical stress gives the same general trend as the experiments.
However, our previous paper found evidence that the smooth isotropic-nematic
transition does not require a supercritical stress~\cite{selinger02}.  In
particular, a smooth transition is seen even in elastomers crosslinked in the
isotropic phase, even under the minimum applied stress required for the
experiment, conditions in which a supercritical stress is unlikely to occur.

An alternative possibility to explain the broadened transition is quenched
disorder in the elastomer.  One specific mechanism to generate quenched disorder
is chemical heterogeneity, which can be represented by random bonds in a lattice
model.  Our simulations show that random-bond disorder can broaden the
isotropic-nematic transition into a smooth crossover, if the magnitude and
length scale of the disorder are large enough.  This numerical result is
consistent with theoretical work on generic random-bond systems by Imry and
Wortis~\cite{imry79}, which argued that weak random-bond disorder should reduce
the first-order discontinuity in a transition, and larger disorder should
eliminate the discontinuity completely.  It is also consistent with other
simulation studies of the isotropic-nematic transition in systems of small
molecules with quenched random impurities~\cite{dadmun92,ilnytskyi99}.  Thus,
random-bond disorder provides a plausible mechanism to explain experimental
results on liquid-crystalline elastomers.

Another type of quenched disorder is heterogeneous local stresses, which can be
represented by random fields in a lattice model.  In general, random fields have
stronger effects on ordered phases and phase transitions than random bonds.  In
our simulations, we find that low random-field disorder broadens the
isotropic-nematic transition, consistent with the experiments, but high
random-field disorder destroys the nematic phase over all temperatures.

Our random-field simulation results are surprising in comparison with
theoretical expectations for random-field systems.  In classic work on quenched
disorder, Imry and Ma~\cite{imry75} showed that arbitrarily small random fields
should destroy long-range order in a \emph{discrete} order parameter (such as
an Ising model) for spatial dimension less than 2, and destroy long-range order
in a \emph{continuous} order parameter for spatial dimension less than 4.  In
our case, the nematic order parameter is continuous, and the spatial dimension
is 3.  Hence, the Imry-Ma argument implies that random fields should destroy
nematic order.  The simulations of high random fields show this effect, but the
simulations of lower random fields show a broad isotropic-nematic transition,
which is a very different effect.

One might ask whether this behavior results from an Imry-Ma domain size that is
large compared with the size of the simulation cell.  To answer that question,
we can estimate the Imry-Ma domain size.  In the Imry-Ma argument, domains form
with a characteristic size $\xi$ such that the boundary energy equals the field
energy.  In our 3D system of continuous directors, the boundary energy is of
order $J \xi$.  To estimate the field energy, recall that we are simulating
blocks of sites with the same random field, as shown in Fig.~3.  Let $b$ be the
linear size of a block.  The field energy for a single block is of order
$\sum_\mathrm{block}{(\bm{h}_i\cdot\bm{n}_i)}^2 \approx h^2 b^3$, and the number
of blocks per domain is of order $(\xi/b)^3$.  As a result, the field energy for
a domain is of order $(h^2 b^3)(\xi/b)^{3/2} \approx h^2 b^{3/2} \xi^{3/2}$.
This argument implies that increasing the block size causes the random field to
be more effective, as is seen in the simulations.  Comparing the boundary energy
with the field energy gives an Imry-Ma domain size of
$\xi \approx J^2/(h^4 b^3)$.  For the simulated values $J=1$, $h=0.4$, and
$b=12$, this size is less than 1 lattice unit, much less than the system size.
Thus, the behavior in our simulations does not arise from a large Imry-Ma domain
size.

We suggest that the new behavior in our simulations arises from the coupling
between the local directors and global elastic strain variable.  The Imry-Ma
prediction is based on an analysis of the energetics of local ordered domains.
This analysis assumes a short-range interaction in the order parameter.  By
contrast, in our model for liquid-crystalline elastomers, the local director at
any lattice site interacts with the global elastic strain, which in turn
interacts with the local director at every other lattice site.  Hence, the
elastic strain mediates an effective long-range interaction between local
directors on different sites.  This changes the assumptions in the Imry-Ma
theory, and hence allows a broad isotropic-nematic transition over a range of
random-field strength.

For a specific numerical test of this suggestion, we perform simulations of a
simplified model \emph{without} the globel elastic strain variable.  The
Hamiltonian for this model consists only of the interaction of
Eq.~(\ref{interaction}) plus the random fields acting on the local directors.
In this case, the interaction is purely short-range, so the Imry-Ma argument
should apply.  Indeed, the simulations show that random fields simply destroy
the nematic order and do not induce a broad isotropic-nematic transition.  This
confirms the concept that a broad transition is a new effect arising from a
strain-mediated long-range interaction.

In a realistic system, the elastic strain-mediated interaction is not
\emph{infinite}-range, as in our model.  However, elastic interactions do have a
power-law form, and hence they can have long-range effects.  In a recent
renormalization-group study, Xing and Radzihovsky~\cite{xing03} have assessed
the effects of elastic interactions on liquid-crystalline elastomers with random
fields.  They find that elastic interactions cause the nematic order to be
robust against the disordering effect of random fields.  Through a power-law
expansion about spatial dimensionality 5, they estimate that nematic order can
be stable down to a critical dimension well below 3.  This is apparently the
same stabilization that we see numerically.  Thus, our simulation shows the
consequence of this stabilization for the temperature-dependent
isotropic-nematic transition.  It is interesting to note that similar
considerations of disorder and long-range elasticity have been seen in models
for magnetic phase transitions in colossal magnetoresistance
materials~\cite{burgy04}.

In conclusion, we have developed a lattice model of the isotropic-nematic
transition in liquid-crystalline elastomers.  The model considers a local
directors coupled to a global elastic distortion variable, and allows both
random-bond and random-field disorder.  Through Monte Carlo simulations of this
model, we find that a uniform elastomer has a mechanical critical point, that
both random-bond disorder and low random-field disorder broaden the
isotropic-nematic transition, and that high random-field disorder destroys the
nematic phase.  The model therefore confirms that the width of the
isotropic-nematic transition can be controlled by heterogeneity in
liquid-crystalline elastomers.

\acknowledgments

We would like to thank M. Warner and T. C. Lubensky for helpful discussions, and
R.~L.~B. Selinger for assistance with the simulation programming.  This work was
supported by the Defense Advanced Research Projects Agency, the Office of Naval
Research, and the Naval Research Laboratory.  Computational resources were
provided by the Maui High Performance Computing Center.

\end{document}